\documentclass[aps,prmaterials,twocolumn,longbibliography,nobibnotes,showkeys]{revtex4-2}

\usepackage{graphicx}
\usepackage{dcolumn}
\usepackage{bm}
\usepackage[version=3]{mhchem}

\begin{document}
\title{Rare-earth defects and defect-related luminescence in ZnS}
\author{Khang Hoang}
\email{khang.hoang@ndsu.edu}
\affiliation{Center for Computationally Assisted Science and Technology \& Department of Physics, North Dakota State University, Fargo, North Dakota 58108, United States}

\date{\today}

\begin{abstract}

Structure and energetics of rare-earth (RE) defects and luminescence of RE and related defects in zincblende zinc sulfide (ZnS) are investigated using hybrid density-functional defect calculations. We find that europium (Eu) is stable predominantly as the divalent Eu$^{2+}$ ion in bulk ZnS. The trivalent Eu$^{3+}$ is structurally and electronically stable, but energetically unfavorable compared to Eu$^{2+}$ due to the presence of low-energy native defects and Eu$^{2+}$-related defect complexes. Other RE dopants, dysprosium (Dy) and erbium (Er), are stable only as Dy$^{3+}$ and Er$^{3+}$, respectively. These results provide an explanation why it is difficult to realize Eu$^{3+}$ in bulk ZnS. A non-negligible Eu$^{3+}$/Eu$^{2+}$ ratio might be achieved with Li co-doping under S-rich (and probably non-equilibrium) synthesis conditions. Optically, Eu-related defects can act as carrier traps for band-to-defect transitions and emit light in the visible range. To assist with experimental optical characterization of the RE defects, we include band-to-defect luminescence involving native defects (Zn vacancies) and/or non-RE impurities (Cu, Cl, and Al) that may also be present in Eu-doped ZnS samples, and assign luminescence centers often observed in experiments to specific defect configurations.

\end{abstract}


\maketitle


\section{Introduction}\label{sec;intro}

ZnS has been of interest for many applications, most notably luminescent devices, display technologies, solar cells, and radiation detection \cite{Shionoya2006,Xu2018AFM,Saleh2019JAP}. More recently, the material has also been identified as a promising host for defect-based qubits \cite{Weber2010PNAS,Gordon2013MRSBull}, mainly due to its wide band gap and weak spin-orbit coupling. Stewart et al.~\cite{Stewart2019OL} reported the observation of single-photon emitters in ZnS nanoparticles and suspected that native zinc vacancy defects could have been the source. In addition to native point defects, rare-earth (RE) impurities, with their $4f$-electron core well shielded by the outer $5s^2$ and $5p^6$ electron shells, can offer interesting features for quantum applications, including quantum memories and quantum computing, as it has already been demonstrated or considered in other RE-doped wide band gap materials \cite{Thiel2011JL,Zhong2019NP,Mitchell2021NP,kinos2021roadmap}. A fundamental understanding of the interaction between the RE dopants and the host material would be essential to identifying RE-related defects suitable for practical applications.  

Experimentally, Eu-doped ZnS has been widely studied, both in the bulk form and in nanomaterials \cite{Hommel1985JCG,Godlewski1986PSS,Ehrhart2008OM,Wang2015Nano,Horoz2016AIPA,Wei2018ACSO}. In the bulk, Eu was reported to be present predominantly as the divalent Eu$^{2+}$ ion, and the trivalent Eu$^{3+}$ was difficult to realize \cite{Hommel1985JCG,Godlewski1986PSS}. Other RE impurities such as Er and Dy, on the other hand, were found to be stable as Er$^{3+}$ and Dy$^{3+}$, respectively \cite{Watts1968PR}. Eu$^{3+}$ and Er$^{3+}$ are of interest for optical applications due to their very sharp intra-$f$ emissions in the visible spectral region. The sharp optical transitions are also correlated with long optical $T_1$ lifetimes which are of interest for quantum applications \cite{Zhong2019NP}. On the theory side, comprehensive first-principles studies of native point defects and non-RE impurities, using a hybrid density-functional theory (DFT)/Hartree-Fock approach, have been reported for the wurtzite \cite{Varley2013APL,Varley2014JAP} and zincblende \cite{Hoang2019CMS} phases of ZnS. Computational studies of RE defects in ZnS and thus a theoretical understanding of their structural, electrical, optical, and thermodynamic properties are, however, currently lacking.

Here, we report an investigation of RE impurities in {\it zincblende} ZnS using hybrid density-functional defect calculations. The zincblende phase is stable at low temperature ($<$1020$^\circ$C) and often reported in the literature \cite{Shionoya2006,Saleh2019JAP}; the wurtzite phase (crystallizes $>$1020$^\circ$C) is expected to have quite similar defect physics (as seen previously in the case of native point defects \cite{Varley2013APL,Hoang2019CMS}). The hybrid DFT/Hartree-Fock approach has been shown to be suitable for the study of defects in RE-doped materials due to its ability to both describe the highly localized RE $4f$ states and reproduce the band gap of the host material \cite{Hoang2021PRM}. 

Specific calculations are carried out mainly for substitutional Eu impurities and defect complexes consisting of Eu and native defects or the Li co-dopant. Eu is chosen for the current study as it is so far most widely studied experimentally. Isolated substitutional Dy and Er impurities are also investigated to illustrate the difference with Eu. On the basis of our results, we discuss the valence of the RE dopants and why it is difficult to realize Eu$^{3+}$ in as-synthesized bulk ZnS, and the role of Eu-related defects as band-to-defect luminescence centers. For comparison and to assist with experimental optical characterization of the RE defects, other luminescence centers involving native defects (Zn vacancies) and/or non-RE (Cu, Cl, and Al) impurities, some of which were previously reported in Ref.~\citenum{Hoang2019CMS}, are also included as they may coexist with the Eu-related centers in actual Eu-doped ZnS samples. Cu, Al, and Cl are chosen because they are often present as intentional or unintentional impurities in ZnS-based materials \cite{Shionoya2006}. 

The goals of this work are thus to help guide defect-controlled synthesis and experimental characterization of defects in ZnS, and to offer specific RE-related defect centers whose properties can be of interest for quantum and optical applications. 

\section{Methodology}\label{sec;method} 

We model defects in ZnS using a supercell approach in which a defect (i.e., a native point defect or an intentional or unintentional impurity) is included in a periodically repeated finite volume of the host material. The formation energy of a defect X in charge state $q$ (with respect to the host lattice) is defined as \cite{Freysoldt2014RMP}     
\begin{align}\label{eq:eform}
E^f({\mathrm{X}}^q)&=&E_{\mathrm{tot}}({\mathrm{X}}^q)-E_{\mathrm{tot}}({\mathrm{bulk}}) -\sum_{i}{n_i\mu_i^\ast} \\ %
\nonumber &&+~q(E_{\mathrm{v}}+\mu_{e})+ \Delta^q ,
\end{align}
where $E_{\mathrm{tot}}(\mathrm{X}^{q})$ and $E_{\mathrm{tot}}(\mathrm{bulk})$ are the total energies of the defect-containing and bulk supercells. $n_{i}$ is the number of atoms of species $i$ that have been added ($n_{i}>0$) or removed ($n_{i}<0$) to form the defect. $\mu_{i}^\ast$ is the atomic chemical potential, representing the energy of the reservoir with which atoms are being exchanged; here, e.g., $\mu_{\rm Zn}^\ast = E_{\rm tot}({\rm Zn}) + \mu_{\rm Zn}$, with $E_{\rm tot}({\rm Zn})$ being the total energy per atom of metallic Zn, and $\mu_{\rm i}^\ast$ is thus referenced to the total energy per atom of $i$ in its elemental phase at 0 K. $\mu_{e}$ is the chemical potential of electrons, i.e., the Fermi level, representing the energy of the electron reservoir, and, as a convention, referenced to the valence-band maximum (VBM) in the bulk ($E_{\mathrm{v}}$). Finally, $\Delta^q$ is the correction term to align the electrostatic potentials of the bulk and defect-containing supercells and to account for finite-size effects on the total energies of charged defects \cite{Freysoldt,Freysoldt11}. In the RE-related defects, for example, $\Delta^q$ is found to be about 0.2--0.4 eV (for $|q|=1$) or 0.4--1.3 eV ($|q|=2$), depending on specific defect configurations.

The formation energy of a defect directly determines the equilibrium concentration \cite{walle:3851}:
\begin{equation}\label{eq;con} 
c=N_{\rm sites}N_{\rm config}\exp{\left(\frac{-E^f}{k_{\rm B}T}\right)}, 
\end{equation} 
where $N_{\rm sites}$ is the number of high-symmetry sites in the lattice (per unit volume) on which the defect can be incorporated, $N_{\rm config}$ is the number of equivalent configurations (per site), and $k_{\rm B}$ is the Boltzmann constant. Note that, when a material is prepared under non-equilibrium conditions, excess defects can be frozen-in and the concentration estimated via Eq.~(\ref{eq;con}) is only the lower bound \cite{Hoang2018JPCM}.

The Fermi level in Eq.~(\ref{eq:eform}) is not a free parameter. Its actual position can be determined by solving the charge-neutrality equation \cite{walle:3851}:
\begin{equation}\label{eq:neutrality}
\sum_{i}c_{i}q_{i}-n_{e}+n_{h}=0,
\end{equation}
where $c_{i}$ and $q_{i}$ are the concentration and charge, respectively, of defect X$_{i}$; $n_{e}$ and $n_{h}$ are free electron and hole concentrations, respectively; and the summation is over all possible defects present in the material.

The {\it thermodynamic} transition level between charge states $q$ and $q'$ of a defect, $\epsilon(q/q')$, is defined as the Fermi-level position at which the formation energy of the defect in charge state $q$ is equal to that in state $q'$ \cite{Freysoldt2014RMP}, i.e.,
\begin{equation}\label{eq;tl}
\epsilon(q/q') = \frac{E^f(X^{q}; \mu_e=0)-E^f(X^{q'}; \mu_e=0)}{q' - q},
\end{equation}
where $E^f(X^{q}; \mu_e=0)$ is the formation energy of the defect X in charge state $q$ when the Fermi level is at the VBM ($\mu_e=0$). This $\epsilon(q/q')$ level [also referred to as the $(q/q')$ level], corresponding to a defect energy level (or, simply, {\it defect level}), would be observed in experiments where the defect in the final charge state $q'$ fully relaxes to its equilibrium configuration after the transition. The {\it optical} transition level $E_{\rm opt}^{q/q'}$ is defined similarly but with the total energy of the final state $q'$ calculated using the lattice configuration of the initial state $q$ \cite{Freysoldt2014RMP}.

The total-energy electronic structure calculations are based on DFT with the Heyd-Scuseria-Ernzerhof (HSE) hybrid functional \cite{heyd:8207}, the projector augmented wave method \cite{PAW1}, and a plane-wave basis set, as implemented in the Vienna {\it Ab Initio} Simulation Package (\textsc{vasp}) \cite{VASP2}. We use the same set of parameters as those in our previous work \cite{Hoang2019CMS}. These include setting the screening length to 10 {\AA} and the Hartree-Fock mixing parameter to 0.32 to match the experimental band gap. Defects are modeled using a 2$\times$2$\times$2 (64-atom) supercell (in which the lattice parameters are fixed to the calculated bulk values but all the internal coordinates are relaxed) and a 2$\times$2$\times$2 Monkhorst-Pack $k$-point mesh for the integrations over the Brillouin zone. The use of the denser, $\Gamma$-centered 3$\times$3$\times$3 $k$-point mesh results in a very small difference in the defect formation energy (e.g., only 1 meV in the case of Eu$_{\rm Zn}^0$). In all the calculations, the plane-wave basis cutoff is 400 eV and spin polarization is included; structural relaxations are performed with the HSE functional and the force threshold is chosen to be 0.02 eV/{\AA}.

The chemical potentials of Zn and S vary over a range determined by the calculated formation enthalpy of ZnS: $\mu_{\rm Zn}+\mu_{\rm S} = \Delta H({\rm  ZnS}) (-1.93$ eV at 0 K) \cite{Hoang2019CMS}. The extreme Zn-rich and S-rich conditions correspond to $\mu_{\rm Zn} = 0$ eV and $\mu_{\rm S} = 0 $ eV, respectively. Specific values for $\mu_{\rm Eu}$, $\mu_{\rm Dy}$, $\mu_{\rm Er}$, and $\mu_{\rm Li}$ are obtained by assuming equilibrium with EuS ($\Delta H = -4.72$ eV at 0 K), Dy$_2$S$_3$ ($-15.10$ eV), ErS ($-4.79$ eV), and Li$_2$S ($-4.40$ eV), respectively. Similarly, the chemical potentials of Cu and Cl are assumed to be limited by the formation of bulk Cu (or CuS, $\Delta H = -0.47$ eV) under the Zn-rich (S-rich) condition and ZnCl$_2$ ($\Delta H = -6.43$ eV), respectively \cite{Hoang2019CMS}. Note that the defect transition levels, $\epsilon(q/q')$ and $E_{\rm opt}^{q/q'}$, are independent of the chemical potentials.

\section{Results and discussion}\label{sec;results}

\begin{figure}
\vspace{0.2cm}
\includegraphics*[width=\linewidth]{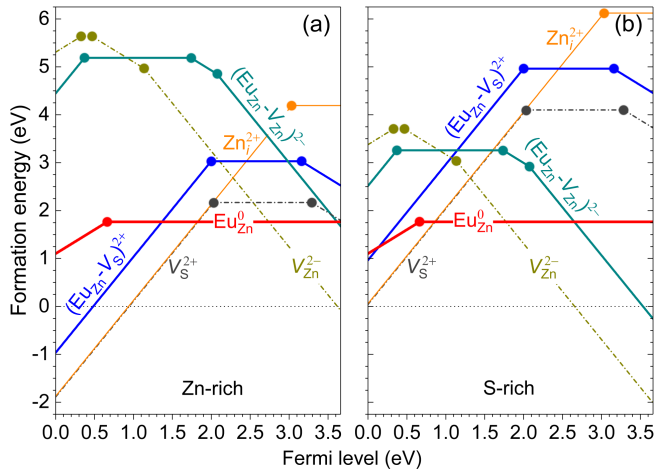}
\caption{Formation energies of Eu-related defects as a function of the Fermi level from the VBM to the CBM, under the extreme (a) Zn-rich and (b) S-rich conditions. The results for native defects $V_{\rm Zn}$, $V_{\rm S}$ and Zn$_i$, previously reported in Ref.~\citenum{Hoang2019CMS}, are also included for reference. For each defect, only segments corresponding to the lowest-energy charge states are shown. The slope of these segments indicates the charge state ($q$): positively (negatively) charged defect configurations have positive (negative) slopes; horizontal segments correspond to neutral conﬁgurations. Large solid dots connecting two segments with different slopes mark the {\it defect levels} [i.e., $\epsilon(q/q')$].} 
\label{fig;fe;eu} 
\end{figure}

We begin by summarizing the basic bulk properties and characteristics of the most relevant native defects. In zincblende ZnS, each Zn is tetrahedrally coordinated with S atoms. The calculated lattice constant is 5.416 {\AA} \cite{Hoang2019CMS} and the Zn--S bond length is 2.345 {\AA}. For comparison, the experimental lattice constant is 5.41 {\AA} \cite{Shionoya2006}. The calculated band gap is 3.66 eV (direct, at $\Gamma$) \cite{Hoang2019CMS}, matching the experimental value (3.7 eV \cite{Shionoya2006}). Zn ($V_{\rm Zn}$) and S ($V_{\rm S}$) vacancies and Zn (Zn$_i$) interstitials are the predominant native point defects \cite{Hoang2019CMS}. In the range of Fermi-level values from the VBM to the conduction-band minimum (CBM), $V_{\rm Zn}$ is found to be stable in the charge states $V_{\rm Zn}^+$ (spin $S=3/2$), $V_{\rm Zn}^0$ ($S=1$), $V_{\rm Zn}^-$ ($S=1/2$), and $V_{\rm Zn}^{2-}$ ($S=0$); $V_{\rm S}$ stable as $V_{\rm S}^{2+}$ ($S=0$), $V_{\rm S}^0$ ($S=0$), and $V_{\rm S}^-$ ($S=1/2$); see Fig.~\ref{fig;fe;eu}. For more details on native defects and non-RE impurities in zincblende ZnS, see Ref.~\citenum{Hoang2019CMS}.         

\subsection{Doping with europium}

Figure \ref{fig;fe;eu} shows the formation energy of Eu-related defects in ZnS. The substitutional Eu impurity (Eu$_{\rm Zn}$) is found to be structurally and electronically stable as Eu$_{\rm Zn}^0$ (i.e., the divalent Eu$^{2+}$ ion at the Zn$^{2+}$ lattice site, with a calculated local spin magnetic moment of $\sim$7 $\mu_{\rm B}$, i.e., spin $S=7/2$) and Eu$_{\rm Zn}^+$ (i.e., the trivalent Eu$^{3+}$ ion at the Zn$^{2+}$ site, with a magnetic moment of $\sim$6 $\mu_{\rm B}$, i.e., $S=3$). The thermodynamic transition level $(+/0)$ of Eu$_{\rm Zn}$ is at 0.66 eV above the VBM; i.e., above (below) this level, Eu$^{2+}$ (Eu$^{3+}$) is energetically more favorable (see further discussion later). In the Eu$_{\rm Zn}^0$ (Eu$_{\rm Zn}^+$) configuration, the Eu--S bond is 2.711--2.712 {\AA} (2.612--2.615 {\AA}), consistent with fact that the ionic radius of Eu$^{2+}$ is larger than that of Eu$^{3+}$. In this work, the valence of an impurity ion in a defect configuration is determined by examining the calculated total and local magnetic moments, electron occupation, and local lattice environment. For example, the calculated local magnetic moment of Eu$_{\rm Zn}^+$ is $\sim$6 $\mu_{\rm B}$ [very close to the total magnetic moment (6 $\mu_{\rm B}$) of the supercell] and almost entirely of $4f$ characters, indicating that there are six unpaired $4f$ electrons localized at the Eu ion, i.e., Eu is electronically stable as the trivalent Eu$^{3+}$ in this defect configuration. The local lattice environment (i.e., the Eu--S bond lengths mentioned earlier) also confirms such determination.   

\begin{figure}
\centering
\includegraphics[width=\linewidth]{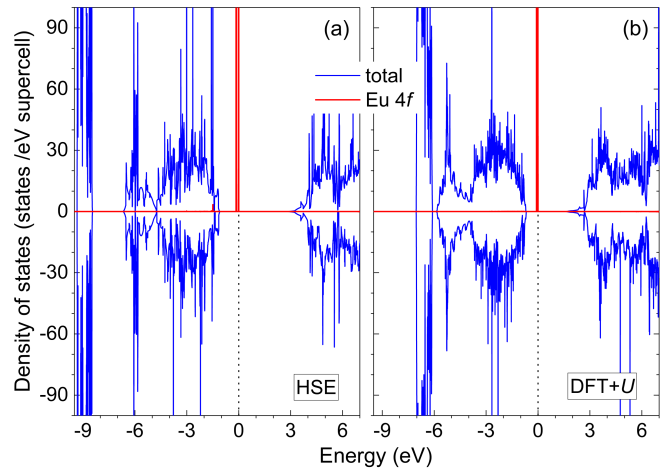}
\caption{Total and Eu $4f$-projected electronic densities of states of Eu-doped ZnS, i.e., the Eu$_{\rm Zn}^0$ defect configuration, obtained in calculations using (a) the HSE functional or (b) the DFT$+$$U$ method ($U^{\rm eff} = 5.0$ eV for Eu $4f$ electrons). The zero of energy is set to the highest occupied state.}
\label{fig;dos}
\end{figure}

\begin{figure*}
\centering
\includegraphics[width=\linewidth]{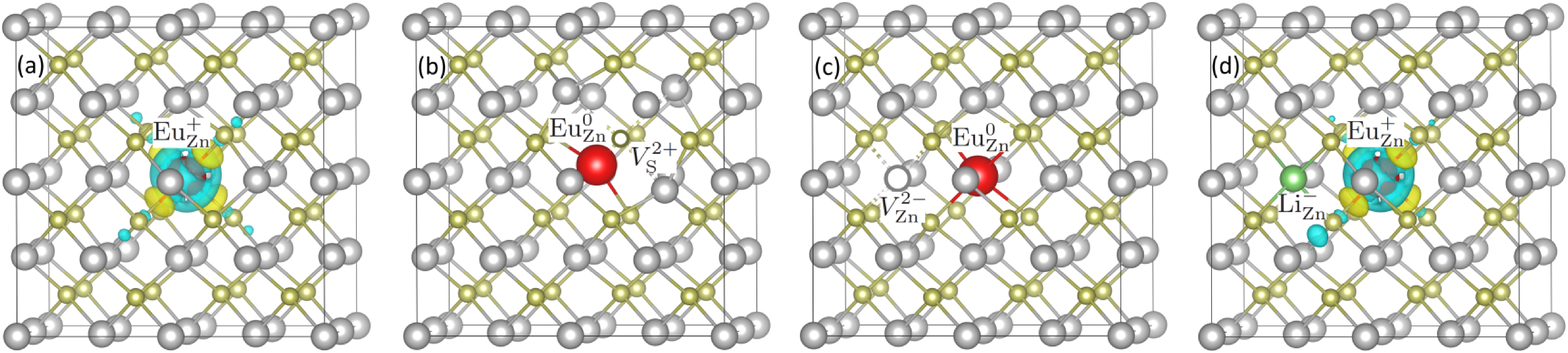}
\caption{Structure of Eu-related defect configurations in ZnS: (a) Eu$_{\rm Zn}^+$, (b) (Eu$_{\rm Zn}$-$V_{\rm S}$)$^{2+}$, (c) (Eu$_{\rm Zn}$-$V_{\rm Zn}$)$^{2-}$, and (d) (Eu$_{\rm Zn}$-Li$_{\rm Zn}$)$^{0}$. In (a) and (d), the associated charge density is included, showing an electron hole highly localized at the Eu ion. The isovalue for the charge-density isosurface is set to 0.02 e/{\AA}$^3$. The large (red) sphere is Eu, medium (gray/green) spheres are Zn/Li, and small (yellow) spheres are N. In (b) and (c), the Zn (S) vacancy is presented by a medium (small) circle.}
\label{fig;struct}
\end{figure*}

The mixed valence of Eu can be understood by examining the electronic structure. Figure \ref{fig;dos} shows the total and Eu $4f$-projected electronic density of states (DOS) of Eu-doped GaN obtained in calculations using HSE \cite{heyd:8207} or DFT$+$$U$ \cite{Dudarev1998}. In both methods, the DFT part is based on the Perdew-Burke-Ernzerhof (PBE) parametrization \cite{GGA} of the generalized gradient approximation. In these calculations, one Zn atom in the 64-atom ZnS supercell is substituted with Eu; i.e., the chemical composition is EuZn$_{31}$S$_{32}$ (and thus the Eu concentration is $\sim$3\%), which corresponds to the Eu$_{\rm Zn}^0$ defect configuration discussed earlier. A $\Gamma$-centered $k$-point mesh of 3$\times$3$\times$3 or denser is used to obtain a good quality DOS. The ground state of Eu$^{2+}$ is $4f^{7}$. In the calculated DOS, we find that Eu introduces seven spin-up occupied $4f$ states in the host band gap. The presence of these in-gap states is key to the stabilization of Eu$^{3+}$ (in addition to Eu$^{2+}$): An electron when removed from the system (i.e., Eu$_{\rm Zn}^0$) will be removed from the highest occupied state, which is, in this case, the highest occupied in-gap Eu $4f$ state. Upon electron removal (or, equivalently, hole creation), Eu$^{2+}$ ($4f^{7}$) becomes Eu$^{3+}$ ($4f^{6}$); or, in the defect notation, Eu$_{\rm Zn}^0$ becomes Eu$_{\rm Zn}^+$. The hole is highly localized at the Eu site, as seen in Fig.~\ref{fig;struct}(a). The main difference between the HSE and DFT$+$$U$ calculations is in the calculated host band gap and hence the position of the Eu $4f$ states with respect to the band edges. 

It should be noted, however, that the Kohn-Sham Eu $4f$ levels as seen in the DOS in Fig.~\ref{fig;dos} must not be directly identified with any defect levels that can be measured in experiments. The experimentally observable defect level $(+/0)$ of Eu$_{\rm Zn}$, for instance, needs to be calculated according to Eq.~(\ref{eq;tl}), as reported in Fig.~\ref{fig;fe;eu} and Table \ref{tab;complex}.   

The Eu$_{\rm Zn}$ defects may not stay isolated but can come close to a native defect and form defect complexes. For that reason, we explore possible association between Eu$_{\rm Zn}$ and $V_{\rm S}$ or $V_{\rm Zn}$. We find that the Eu$_{\rm Zn}$-$V_{\rm S}$ complex introduces two defect levels in the host band gap: $(2+/0)$ at 2.00 eV above the VBM and $(0/-)$ at 0.50 eV below the CBM; see Fig.~\ref{fig;fe;eu}. Eu is stable as Eu$^{2+}$ in all the stable configurations of the complex, and (Eu$_{\rm Zn}$-$V_{\rm S}$)$^q$ is a complex of Eu$_{\rm Zn}^0$ and $V_{\rm S}^{2+}$ [for $q=2+$; see Fig.~\ref{fig;struct}(b)], $V_{\rm S}^{0}$ ($q=0$), or $V_{\rm S}^{-}$ ($q=-$). In these defect configurations, the Eu ion is off-center by 0.70 {\AA} ($q=2+$), 0.45 {\AA} ($q=0$), or 0.50 {\AA} ($q=-$) toward the S vacancy. The complex between Eu$_{\rm Zn}^+$ and $V_{\rm S}^+$ is energetically not stable, mainly due to the repulsive Coulomb interaction. The Eu$_{\rm Zn}$-$V_{\rm Zn}$ complex, on the other hand, has three defect levels in the host band gap: $(2+/0)$ at 0.37 eV, $(0/-)$ at 1.74 eV, and $(-/2-)$ at 2.07 eV above the VBM; see Fig.~\ref{fig;fe;eu}. Eu can be stable as Eu$^{2+}$ or Eu$^{3+}$ in the complex, depending on the complex's specific charge state. (Eu$_{\rm Zn}$-$V_{\rm Zn}$)$^q$ is a complex of Eu$_{\rm Zn}^+$ and $V_{\rm Zn}^{+}$ (for $q=2+$) or $V_{\rm Zn}^{2-}$ ($q=-$), or a complex of Eu$_{\rm Zn}^0$ and $V_{\rm Zn}^{0}$ ($q=0$) or $V_{\rm Zn}^{2-}$ [$q=2-$; see Fig.~\ref{fig;struct}(c)]. The stability of (Eu$_{\rm Zn}$-$V_{\rm Zn}$)$^{2+}$ (and hence of Eu$^{3+}$ in the complex) indicates that the local elastic interaction also plays an important role in defect association. Indeed, we find that Eu is off-center by 2.12 {\AA} in that defect. In the other charge states, the Eu is off-center by 1.85 {\AA} ($q=0$), 1.95 {\AA} ($q=-$), or 1.94 {\AA} ($q=2-$) toward the Zn vacancy. In all the Eu$_{\rm Zn}$-$V_{\rm S}$ and Eu$_{\rm Zn}$-$V_{\rm Zn}$ configurations, in addition to the mentioned off-centering of the Eu ion, there is also relaxation of the neighboring Zn and/or S atoms; however, the effect is limited mainly to the nearest neighbors. 

\begin{table*}
\caption{Rare-earth (RE) related defects in ZnS: The stable valence state of the RE ion, constituent defects, binding energy ($E_b$, with respect to the isolated constituents), spin magnetic moment ($M$, per supercell), and defect levels [$\epsilon(q/q')$, with respect to the VBM ($E_{\rm v}$) or the CBM ($E_{\rm c}$)]. Spin-polarized defects in a defect complex are found to interact ferromagnetically; the magnetic moment (and hence the spin) of the complex is thus equal to the sum of those of the isolated constituent defects, except in the case of (Eu$_{\rm Zn}$-$V_{\rm Zn}$)$^q$ with $q=2+, 0$ (see the text).}\label{tab;complex}
\begin{center}
\begin{ruledtabular}
\begin{tabular}{lllrcl}
Defect &RE ion &Constituents & $E_{b}$ (eV) & $M$ ($\mu_{\rm B}$) & Defect levels (eV) \\
\colrule
Eu$_{\rm Zn}^+$&Eu$^{3+}$&Eu$_{\rm Zn}^+$&&6&$\epsilon(+/0)=E_{\rm v}+0.66$\\
Eu$_{\rm Zn}^0$&Eu$^{2+}$&Eu$_{\rm Zn}^0$&&7&\\
(Eu$_{\rm Zn}$-$V_{\rm S}$)$^{2+}$ &Eu$^{2+}$& Eu$_{\rm Zn}^0$ + $V_{\rm S}^{2+}$ &0.84&7&$\epsilon(2+/0)=E_{\rm v}+2.00$\\
(Eu$_{\rm Zn}$-$V_{\rm S}$)$^{0}$ &Eu$^{2+}$& Eu$_{\rm Zn}^0$ + $V_{\rm S}^{0}$ &0.90&7&$\epsilon(0/-)=E_{\rm c}-0.50$\\
(Eu$_{\rm Zn}$-$V_{\rm S}$)$^{-}$ &Eu$^{2+}$& Eu$_{\rm Zn}^0$ + $V_{\rm S}^{-}$ &1.04&8&\\
(Eu$_{\rm Zn}$-$V_{\rm Zn}$)$^{2+}$ &Eu$^{3+}$&Eu$_{\rm Zn}^+$ + $V_{\rm Zn}^+$&1.96&7&$\epsilon(2+/0)=E_{\rm v}+0.37$\\
(Eu$_{\rm Zn}$-$V_{\rm Zn}$)$^{0}$ &Eu$^{2+}$&Eu$_{\rm Zn}^0$ + $V_{\rm Zn}^0$&2.21&7&$\epsilon(0/-)=E_{\rm v}+1.74$\\
(Eu$_{\rm Zn}$-$V_{\rm Zn}$)$^{-}$ &Eu$^{3+}$& Eu$_{\rm Zn}^+$ + $V_{\rm Zn}^{2-}$ &1.42&6&$\epsilon(-/2-)=E_{\rm v}+2.07$\\
(Eu$_{\rm Zn}$-$V_{\rm Zn}$)$^{2-}$ &Eu$^{2+}$& Eu$_{\rm Zn}^0$ + $V_{\rm Zn}^{2-}$ &0.01&7&\\
(Eu$_{\rm Zn}$-Li$_{\rm Zn}$)$^{0}$ &Eu$^{3+}$& Eu$_{\rm Zn}^+$ + Li$_{\rm Zn}^-$ &0.50&6&$\epsilon(0/-)=E_{\rm v}+1.27$\\
(Eu$_{\rm Zn}$-Li$_{\rm Zn}$)$^{-}$ &Eu$^{2+}$& Eu$_{\rm Zn}^0$ + Li$_{\rm Zn}^-$ &$-$0.11&7&\\
Dy$_{\rm Zn}^+$&Dy$^{3+}$&Dy$_{\rm Zn}^+$&&5&\\
Er$_{\rm Zn}^+$&Er$^{3+}$&Er$_{\rm Zn}^+$&&3&\\
\end{tabular}
\end{ruledtabular}
\end{center}
\begin{flushleft}
\end{flushleft}
\end{table*}

Table \ref{tab;complex} lists the characteristics of the defect complexes, including the calculated binding energy. The binding energy is defined as the difference between the sum of the formation energies of the constituent defects and that of the complex. The identity of the constituent defects in a defect complex configuration is determined by examining the calculated total and local magnetic moments, electron occupation, and local lattice environment. Possible combinations of constituent defects that can make up a complex are limited by the availability of their respective structurally and electronically stable defect configurations \cite{Hoang2021PRM}. In Table \ref{tab;complex}, the spin of a defect complex is the sum of those of the isolated constituent defects, except in the case of (Eu$_{\rm Zn}$-$V_{\rm Zn}$)$^{q}$ with $q=2+$ or $0$ where there is a deviation from the expected value, probably due to the strong local lattice relaxations according to which the spins of the constituent defects may not be conserved. Also note that (Eu$_{\rm Zn}$-$V_{\rm Zn}$)$^{2-}$ has a very low calculated binding energy (within the error bar of the calculation, which is rather large for this particular defect configuration due to uncertainties, caused by large lattice relaxations, in determining the potential alignment term) and thus may not be stable as a complex in Eu-doped ZnS samples synthesized under thermodynamic equilibrium.

Considering only the Eu-related defects, we find that the lowest-energy configuration under the Zn-rich condition is (Eu$_{\rm Zn}$-$V_{\rm S}$)$^{2+}$ ($0$ eV $<$ $\mu_e$ $<$ $1.35$ eV), Eu$_{\rm Zn}^0$ ($1.35$ eV $<$ $\mu_e$ $<$ $3.60$ eV), or (Eu$_{\rm Zn}$-$V_{\rm Zn}$)$^{2-}$ ($3.60$ eV $<$ $\mu_e$ $<$ $3.66$ eV); see Fig.~\ref{fig;fe;eu}(a). These defects all have Eu$^{2+}$; i.e., Eu is energetically favorable only as Eu$^{2+}$ in all these configurations. Under the S-rich condition, the lowest-energy defect is (Eu$_{\rm Zn}$-$V_{\rm S}$)$^{2+}$ (for $0$ eV $<$ $\mu_e$ $<$ $0.15$ eV), Eu$_{\rm Zn}^+$ ($0.15$ eV $<$ $\mu_e$ $<$ $0.65$ eV), Eu$_{\rm Zn}^0$ ($0.65$ eV $<$ $\mu_e$ $<$ $2.65$ eV), or (Eu$_{\rm Zn}$-$V_{\rm Zn}$)$^{2-}$ ($2.65$ eV $<$ $\mu_e$ $<$ $3.66$ eV); see Fig.~\ref{fig;fe;eu}(b). The trivalent Eu$^{3+}$, i.e., Eu$_{\rm Zn}^+$, can thus be energetically favorable under this condition, but only in a small range of Fermi-level values near the VBM. 

To see if that Eu$^{3+}$-favorable range is accessible, however, we need to take other defects that may be present in the material into consideration as well. Low-energy, positively charged native point defects, such as $V_{\rm S}^{2+}$ and Zn$_i^{2+}$ \cite{Hoang2019CMS}, can further reduce the Eu$^{3+}$-favorable range or make it inaccessible. Indeed, as shown in Fig.~\ref{fig;fe;eu}(b), the Fermi level of the system under the extreme S-rich condition is predominantly determined by $V_{\rm Zn}^{2-}$, $V_{\rm S}^{2+}$, and Zn$_i^{2+}$ (i.e., the lowest-energy charged defects) at which Eu$_{\rm Zn}^0$ (i.e., Eu$^{2+}$, {\it not} Eu$^{3+}$) is energetically most favorable.

\subsection{(Co-)doping with lithium} 

\begin{figure}
\vspace{0.2cm}
\includegraphics*[width=\linewidth]{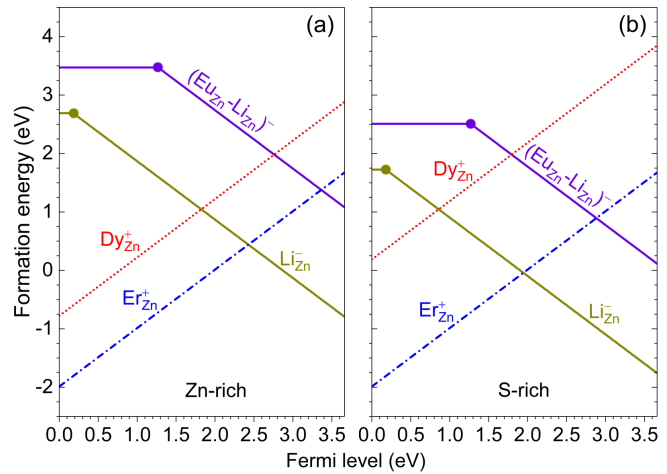}
\caption{Formation energies of Li-related, Dy, and Er defects as a function of the Fermi level from the VBM to the CBM, under the (a) Zn-rich and (b) S-rich conditions. Only segments corresponding to the lowest-energy charge states are shown.}
\label{fig;fe;others} 
\end{figure}

In order to enhance the stability of Eu$^{3+}$, one may co-dope the material with acceptor-like defects such as Li$_{\rm Zn}$, preferably under the S-rich condition to minimize charge compensation caused by positively charged native defects (specifically, $V_{\rm S}^{2+}$ and Zn$_i^{2+}$) \cite{Hoang2019CMS}; see also Fig.~\ref{fig;fe;eu}. We find that Li$_{\rm Zn}$ has a thermodynamic transition level, $(0/-)$, at 0.18 eV above the VBM; see Fig.~\ref{fig;fe;others}. The Eu$_{\rm Zn}$-Li$_{\rm Zn}$ complex introduces a defect level that is higher in energy: $(0/-)$ at 1.27 eV above the VBM. (Eu$_{\rm Zn}$-Li$_{\rm Zn}$)$^0$ is a complex of Eu$_{\rm Zn}^+$ and Li$_{\rm Zn}^-$, whereas (Eu$_{\rm Zn}$-Li$_{\rm Zn}$)$^-$ is a complex of Eu$_{\rm Zn}^0$ and Li$_{\rm Zn}^-$; see Table \ref{tab;complex} for more details. Figure \ref{fig;struct}(d) shows the structure of (Eu$_{\rm Zn}$-Li$_{\rm Zn}$)$^0$; the hole is highly localized at the Eu site. Note that, given its low (or even negative) calculated binding energy, Eu$_{\rm Zn}$-Li$_{\rm Zn}$ is unlikely to be stable as a defect complex in (Eu,Li)-doped ZnS samples under thermodynamic equilibrium.

The Li co-doping can, {\it in principle}, affect the stability of Eu$^{3+}$ in two aspects. One, Li$_{\rm Zn}^-$ may shift the Fermi level toward the VBM as the system reestablishes charge neutrality, thus potentially placing the Fermi level in the Eu$^{3+}$-favorable region (i.e., $\mu_e \le 0.66$ eV; see Fig.~\ref{fig;fe;eu}). Given the choice of the chemical potentials, see Sec.~\ref{sec;method}, we find that the Fermi level of the system co-doped with Li is at 1.64 eV (0.73 eV) under the Zn-rich (S-rich) condition, determined predominantly by Li$_{\rm Zn}^-$, $V_{\rm S}^{2+}$, and Zn$_i^{2+}$ (and, in the S-rich case, some contribution from free electron holes). The Fermi-level position is determined by solving Eq.~(\ref{eq:neutrality}) at 1250$^\circ$C (the temperature used in some experiments) \cite{Hommel1985JCG} for the native defects and Li defects using the \textsc{sc-fermi} code \cite{Buckeridge2019CPC}. Co-doping with Li under the extreme S-rich condition may thus place the Fermi level near the ($+/0$) level of Eu$_{\rm Zn}$. However, given that the error bar of defect calculations is typically about 0.1--0.2 eV, it is not clear at this point if a significant Eu$^{3+}$/Eu$^{2+}$ ratio can be realized in bulk ZnS under actual synthesis conditions. Two, defect association between Li$_{\rm Zn}^-$ and Eu$_{\rm Zn}^+$ can lead to the formation of (Eu$_{\rm Zn}$-Li$_{\rm Zn}$)$^0$ which has a larger Eu$^{3+}$-favorable range than the isolated Eu$_{\rm Zn}^+$ (1.27 eV vs.~0.66 eV). However, given the binding energy reported above, the complex is unlikely to form, except probably under certain non-equilibrium conditions. Note that the former is a {\it global} effect (i.e., the Li and Eu dopants do not need to form a defect complex), whereas the latter, if occurs, is {\it local} (which involves direct defect--defect interaction). For a more detailed discussion of similar global and local effects on the valence of RE ion in the case of Eu-doped GaN, see Ref.~\citenum{Hoang2021PRM}.

Experimentally, S-rich environments (e.g., under CS$_2$ flow) and/or Li co-doping have been employed in attempts to stabilize Eu$^{3+}$; yet the presence of the trivalent ion in bulk ZnS has not been established \cite{Hommel1985JCG,Godlewski1986PSS}. Our results explain why it is difficult to realize Eu$^{3+}$. Further attempts, with Li co-doping under the extreme S-rich condition like discussed earlier, may be needed. A measurement of the Fermi level would also be helpful as its position with respect to the thermodynamic transition level ($+/0$) of Eu$_{\rm Zn}$ determines the Eu$^{3+}$/Eu$^{2+}$ ratio. Note that the incorporation of Eu$^{3+}$ into ZnS nanoparticles apparently can be achieved with more ease \cite{Ehrhart2008OM,Wang2015Nano,Horoz2016AIPA,Wei2018ACSO}.

\subsection{Other rare-earth impurities}

To illustrate the difference with Eu, we also investigate other RE dopants such Dy and Er and find that they are stable only as Dy$^{3+}$ and Er$^{3+}$ in ZnS, which is in agreement with experiments \cite{Watts1968PR,Hommel1985JCG}. This can be seen in Fig.~\ref{fig;fe;others} where the Dy$_{\rm Zn}$ and Er$_{\rm Zn}$ defects are stable only as Dy$_{\rm Zn}^+$ (spin $S=5/2$) and Er$_{\rm Zn}^+$ ($S=3/2$), respectively. As positively charged defects, they may form defect complexes with native point defects and other impurities, especially negatively charged ones. This requires further investigation and is not discussed here. A comprehensive study of all RE dopants is beyond the scope of the current work. 

\begin{figure}[t]%
\vspace{0.2cm}
\includegraphics*[width=\linewidth]{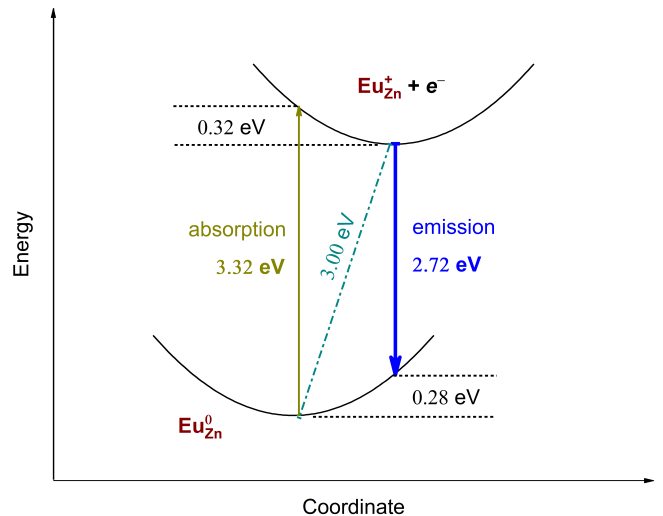}
\caption{Configuration-coordinate diagram illustrating optical emission (down arrow) and absorption (up arrow) processes, $E_{\rm opt}^{q/q'}$, involving Eu$_{\rm Zn}$ in ZnS. The thermal energy [also called the zero-phonon line (ZPL), the dash-dotted line] is the thermodynamic transition level $\epsilon(q/q')$ relative to CBM. The values sandwiched between two dotted lines are the relaxation energies (the Franck-Condon shifts). Axes are not to scale.}
\label{fig;cc} 
\end{figure}

\begin{table*}
\caption{Defect-related luminescence in ZnS: The emission process involves capturing an electron ($e^-$) from the CBM (or some shallow donor level) or a hole ($h^+$) from the VBM (or some shallow acceptor level) by a defect configuration.}\label{tab;opt}
\begin{center}
\begin{ruledtabular}
\begin{tabular}{lcccc}
Optical transition & Emission peak (eV) & ZPL (eV) & Reference & Center acronym\\
\colrule
Eu$_{\rm Zn}^{+}$ + $e^-$ $\rightarrow$ Eu$_{\rm Zn}^{0}$ & 2.72 & 3.00 & This work \\
(Eu$_{\rm Zn}$-$V_{\rm S}$)$^{-}$  + $h^+$ $\rightarrow$ (Eu$_{\rm Zn}$-$V_{\rm S}$)$^{0}$ & 2.68 & 3.16 & This work \\
(Eu$_{\rm Zn}$-$V_{\rm Zn}$)$^{+}$  + $e^-$ $\rightarrow$ (Eu$_{\rm Zn}$-$V_{\rm Zn}$)$^{0}$ & 2.10 & 3.39 & This work \\
(Eu$_{\rm Zn}$-Li$_{\rm Zn}$)$^{0}$  + $e^-$ $\rightarrow$ (Eu$_{\rm Zn}$-Li$_{\rm Zn}$)$^{-}$ & 2.10 & 2.39 & This work \\
$V_{\rm Zn}^{-}$ + $e^-$ $\rightarrow$ $V_{\rm Zn}^{2-}$ & 2.16 & 2.52 & This work \\
($V_{\rm Zn}$-Cl$_{\rm S}$)$^{0}$ + $e^-$ $\rightarrow$ ($V_{\rm Zn}$-Cl$_{\rm S}$)$^{-}$ & 2.54 & 2.95 & This work & SA \\ 
Cu$_{\rm Zn}^{0}$ + $e^-$ $\rightarrow$ Cu$_{\rm Zn}^{-}$ & 2.28 & 2.44 & Ref.~\citenum{Hoang2019CMS} & G-Cu \\ 
(Cu$_{\rm Zn}$-Cl$_i$)$^{+}$  + $e^-$ $\rightarrow$ (Cu$_{\rm Zn}$-Cl$_i$)$^{0}$ & 2.80 & 3.00 & This work & B-Cu \\ 
(Cu$_{\rm Zn}$-Cl$_{\rm S}$)$^{+}$  + $e^-$ $\rightarrow$ (Cu$_{\rm Zn}$-Cl$_{\rm S}$)$^{0}$ & 2.81 & 3.01 & Ref.~\citenum{Hoang2019CMS}\\
(Cu$_{\rm Zn}$-Al$_{\rm Zn}$)$^{+}$  + $e^-$ $\rightarrow$ (Cu$_{\rm Zn}$-Al$_{\rm Zn}$)$^{0}$ & 2.79 & 2.93 & Ref.~\citenum{Hoang2019CMS} \\
\end{tabular}
\end{ruledtabular}
\end{center}
\begin{flushleft}
\end{flushleft}
\end{table*}

\subsection{Band-to-defect luminescence}

Like native defects and non-RE impurities with defect energy levels in the band gap of a semiconductor host, the Eu-related defects can also act as carrier traps in band-to-defect optical transitions \cite{Hoang2021PRM}. Under illumination, e.g., the isolated Eu$_{\rm Zn}^0$ can absorb a photon and become ionized (i.e., Eu$_{\rm Zn}^+$) with the removed electron being excited into the conduction band. Eu$_{\rm Zn}^+$ can then capture an electron from the CBM, e.g., previously excited from Eu$_{\rm Zn}^0$ to the conduction band, and emit a photon. Figure \ref{fig;cc} illustrates these absorption and emission processes. The peak emission energy corresponding to the optical transition level $E_{\rm opt}^{+/0}$, i.e., the energy difference between Eu$_{\rm Zn}^0$ and Eu$_{\rm Zn}^+$ in the lattice configuration of Eu$_{\rm Zn}^0$, is calculated to be 2.72 eV (blue), with a relaxation energy of 0.28 eV. Similar emission processes involving other Eu-related defect configurations such as (Eu$_{\rm Zn}$-$V_{\rm S}$)$^-$, (Eu$_{\rm Zn}$-$V_{\rm Zn}$)$^{+}$, and (Eu$_{\rm Zn}$-Li$_{\rm Zn}$)$^{0}$ are listed in Table \ref{tab;opt}. 

Given the defect landscape discussed earlier as well as that reported in Ref.~\citenum{Hoang2019CMS}, Eu$_{\rm Zn}$ is unlikely the only luminescence center in Eu-doped ZnS samples. For example, we find that $V_{\rm Zn}$ can be the source of a yellow/green emission. The $V_{\rm Zn}^{-}$ + $e^-$ $\rightarrow$ $V_{\rm Zn}^{2-}$ transition has an emission peak at 2.16 eV, which is in excellent agreement with the experimental value (2.18 eV) reported by Lee et al.~\cite{Lee1982SSC}. The thermodynamic transition level $(-/2-)$ of $V_{\rm Zn}$ is at 1.14 eV above the VBM \cite{Hoang2019CMS} and, in going from $V_{\rm Zn}^{2-}$ (spin $S=0$) to $V_{\rm Zn}^{-}$ ($S=1/2$), the hole is localized at one of neighboring S sites of the vacancy, which is also consistent with the reported experimental study \cite{Lee1982SSC}. As mentioned earlier, $V_{\rm Zn}$ was suspected to the source of the single-photon emitters observed in ZnS nanoparticles \cite{Stewart2019OL}.

To assist with experimental optical characterization of the RE defects, we also include in Table \ref{tab;opt} emission processes related to non-RE impurities. As mentioned in Sec.~\ref{sec;intro}, Cu, Cl, and Al are often incorporated intentionally or present unintentionally in ZnS phosphors. The defect complex consisting of $V_{\rm Zn}$ and the substitutional Cl impurity (Cl$_{\rm S}$), i.e., $V_{\rm Zn}$-Cl$_{\rm S}$, is regarded as the ``SA'' (SA=self-activated) luminescence center in the literature \cite{Shionoya2006,Saleh2019JAP}. We find that the complex introduces two defect levels in the host band gap: $(+/0)$ at 0.17 eV and $(0/-)$ at 0.71 eV above the VBM. The ($V_{\rm Zn}$-Cl$_{\rm S}$)$^{0}$ + $e^-$ $\rightarrow$ ($V_{\rm Zn}$-Cl$_{\rm S}$)$^{-}$ transition results in a peak emission energy at 2.54 eV, in agreement with the values (2.58--2.64 eV) reported for the SA center \cite{Saleh2019JAP}. The association between the substitutional Cu impurity (Cu$_{\rm Zn}$) and the Cl interstitial (Cl$_i$), i.e., Cu$_{\rm Zn}$-Cl$_i$, is often denoted as the ``B-Cu'' (B=blue emission) center \cite{Shionoya2006,Saleh2019JAP}. This defect complex has two defect levels: $(+/0)$ at 0.66 eV above the VBM and $(0/2-)$ at 0.67 eV below the CBM. The (Cu$_{\rm Zn}$-Cl$_i$)$^{+}$  + $e^-$ $\rightarrow$ (Cu$_{\rm Zn}$-Cl$_i$)$^{0}$ transition leads to an emission peak of 2.80 eV, which is close to the values (2.92--2.99 eV) reported for the B-Cu center \cite{Saleh2019JAP}. Finally, as reported in Ref.~\citenum{Hoang2019CMS}, the isolated Cu$_{\rm Zn}$ is a source of green luminescence (2.28 eV) in Cu-doped ZnS, and can be identified with the ``G-Cu'' (G=green emission) center \cite{Shionoya2006,Saleh2019JAP}; the Cu$_{\rm Zn}$-Al$_{\rm Zn}$ (2.79 eV) and Cu$_{\rm Zn}$-Cl$_{\rm S}$ (2.81 eV) complexes are sources of blue luminescence. More details of the Cu, Al, and Cl-related defects can be found in Ref.~\citenum{Hoang2019CMS}. 

\section{Conclusions} 

We have carried out an investigation of RE and other relevant defects in ZnS using hybrid density-functional calculations. We find europium (Eu) is stable predominantly as the divalent Eu$^{2+}$ ion in bulk ZnS. The trivalent Eu$^{3+}$ is stable structurally and electronically, as evidenced in the calculated atomic structure, formation energy, and electronic structure of the isolated Eu$_{\rm Zn}^0$ defect; however, it is energetically unfavorable compared to Eu$^{2+}$ due to the presence of low-energy native defects and Eu$^{2+}$-related defect complexes which prevents the system from accessing the range of Fermi-level values in which Eu$^{3+}$ is favorable. The other RE dopants, Dy and Er, are stable only as Dy$^{3+}$ and Er$^{3+}$, respectively. The results are consistent with experimental data on the valence of the RE dopants and provide an explanation why it is difficult to stabilize Eu$^{3+}$ in bulk ZnS. We also suggest that a non-negligible Eu$^{3+}$/Eu$^{2+}$ ratio {\it might} be achieved with Li co-doping under the extreme S-rich (and probably non-equilibrium) synthesis condition. Regarding the optical properties, we find that Eu-related defects can act as carrier traps for band-to-defect optical transitions and emit light in the visible range. Results for band-to-defect luminescence involving native defects ($V_{\rm Zn}$) and/or non-RE impurities (Cu, Cl, Al) are also reported and luminescence centers often observed in experiments, e.g., SA, G-Cu, and B-Cu, are identified.

\begin{acknowledgments}

This work used resources of the Center for Computationally Assisted Science and Technology (CCAST) at North Dakota State University, which were made possible in part by NSF MRI Award No.~2019077. 

\end{acknowledgments}


\begin{thebibliography}{33}%
\makeatletter
\providecommand \@ifxundefined [1]{%
 \@ifx{#1\undefined}
}%
\providecommand \@ifnum [1]{%
 \ifnum #1\expandafter \@firstoftwo
 \else \expandafter \@secondoftwo
 \fi
}%
\providecommand \@ifx [1]{%
 \ifx #1\expandafter \@firstoftwo
 \else \expandafter \@secondoftwo
 \fi
}%
\providecommand \natexlab [1]{#1}%
\providecommand \enquote  [1]{``#1''}%
\providecommand \bibnamefont  [1]{#1}%
\providecommand \bibfnamefont [1]{#1}%
\providecommand \citenamefont [1]{#1}%
\providecommand \href@noop [0]{\@secondoftwo}%
\providecommand \href [0]{\begingroup \@sanitize@url \@href}%
\providecommand \@href[1]{\@@startlink{#1}\@@href}%
\providecommand \@@href[1]{\endgroup#1\@@endlink}%
\providecommand \@sanitize@url [0]{\catcode `\\12\catcode `\$12\catcode
  `\&12\catcode `\#12\catcode `\^12\catcode `\_12\catcode `\%12\relax}%
\providecommand \@@startlink[1]{}%
\providecommand \@@endlink[0]{}%
\providecommand \url  [0]{\begingroup\@sanitize@url \@url }%
\providecommand \@url [1]{\endgroup\@href {#1}{\urlprefix }}%
\providecommand \urlprefix  [0]{URL }%
\providecommand \Eprint [0]{\href }%
\providecommand \doibase [0]{https://doi.org/}%
\providecommand \selectlanguage [0]{\@gobble}%
\providecommand \bibinfo  [0]{\@secondoftwo}%
\providecommand \bibfield  [0]{\@secondoftwo}%
\providecommand \translation [1]{[#1]}%
\providecommand \BibitemOpen [0]{}%
\providecommand \bibitemStop [0]{}%
\providecommand \bibitemNoStop [0]{.\EOS\space}%
\providecommand \EOS [0]{\spacefactor3000\relax}%
\providecommand \BibitemShut  [1]{\csname bibitem#1\endcsname}%
\let\auto@bib@innerbib\@empty
\bibitem [{\citenamefont {Shionoya}(2006)}]{Shionoya2006}%
  \BibitemOpen
  \bibfield  {author} {\bibinfo {author} {\bibfnamefont {S.}~\bibnamefont
  {Shionoya}},\ }\bibfield  {title} {\enquote {\bibinfo {title} {Principal
  phosphor materials and their optical properties},}\ }in\ \href
  {https://books.google.com/books?id=I9O1K20-uo4C} {\emph {\bibinfo {booktitle}
  {Phosphor Handbook}}},\ \bibinfo {series and number} {CRC Press Laser and
  Optical Science and Technology Series},\ \bibinfo {editor} {edited by\
  \bibinfo {editor} {\bibfnamefont {S.}~\bibnamefont {Shionoya}}, \bibinfo
  {editor} {\bibfnamefont {W.~M.}\ \bibnamefont {Yen}},\ and\ \bibinfo {editor}
  {\bibfnamefont {H.}~\bibnamefont {Yamamoto}}}\ (\bibinfo  {publisher} {CRC
  Press},\ \bibinfo {year} {2006})\ pp.\ \bibinfo {pages}
  {247--274}\BibitemShut {NoStop}%
\bibitem [{\citenamefont {Xu}\ \emph {et~al.}(2018)\citenamefont {Xu},
  \citenamefont {Li}, \citenamefont {Chen}, \citenamefont {Cai}, \citenamefont
  {Long},\ and\ \citenamefont {Fang}}]{Xu2018AFM}%
  \BibitemOpen
  \bibfield  {author} {\bibinfo {author} {\bibfnamefont {X.}~\bibnamefont
  {Xu}}, \bibinfo {author} {\bibfnamefont {S.}~\bibnamefont {Li}}, \bibinfo
  {author} {\bibfnamefont {J.}~\bibnamefont {Chen}}, \bibinfo {author}
  {\bibfnamefont {S.}~\bibnamefont {Cai}}, \bibinfo {author} {\bibfnamefont
  {Z.}~\bibnamefont {Long}},\ and\ \bibinfo {author} {\bibfnamefont
  {X.}~\bibnamefont {Fang}},\ }\bibfield  {title} {\enquote {\bibinfo {title}
  {{Design Principles and Material Engineering of ZnS for Optoelectronic
  Devices and Catalysis}},}\ }\href {https://doi.org/10.1002/adfm.201802029}
  {\bibfield  {journal} {\bibinfo  {journal} {Adv. Funct. Mater.}\ }\textbf
  {\bibinfo {volume} {28}},\ \bibinfo {pages} {1802029} (\bibinfo {year}
  {2018})}\BibitemShut {NoStop}%
\bibitem [{\citenamefont {Saleh}\ \emph {et~al.}(2019)\citenamefont {Saleh},
  \citenamefont {Lynn}, \citenamefont {Jacobsohn},\ and\ \citenamefont
  {McCloy}}]{Saleh2019JAP}%
  \BibitemOpen
  \bibfield  {author} {\bibinfo {author} {\bibfnamefont {M.}~\bibnamefont
  {Saleh}}, \bibinfo {author} {\bibfnamefont {K.~G.}\ \bibnamefont {Lynn}},
  \bibinfo {author} {\bibfnamefont {L.~G.}\ \bibnamefont {Jacobsohn}},\ and\
  \bibinfo {author} {\bibfnamefont {J.~S.}\ \bibnamefont {McCloy}},\ }\bibfield
   {title} {\enquote {\bibinfo {title} {Luminescence of undoped commercial
  {ZnS} crystals: {A} critical review and new evidence on the role of
  impurities using photoluminescence and electrical transient spectroscopy},}\
  }\href {https://doi.org/10.1063/1.5084738} {\bibfield  {journal} {\bibinfo
  {journal} {J. Appl. Phys.}\ }\textbf {\bibinfo {volume} {125}},\ \bibinfo
  {pages} {075702} (\bibinfo {year} {2019})}\BibitemShut {NoStop}%
\bibitem [{\citenamefont {Weber}\ \emph {et~al.}(2010)\citenamefont {Weber},
  \citenamefont {Koehl}, \citenamefont {Varley}, \citenamefont {Janotti},
  \citenamefont {Buckley}, \citenamefont {Van~de Walle},\ and\ \citenamefont
  {Awschalom}}]{Weber2010PNAS}%
  \BibitemOpen
  \bibfield  {author} {\bibinfo {author} {\bibfnamefont {J.~R.}\ \bibnamefont
  {Weber}}, \bibinfo {author} {\bibfnamefont {W.~F.}\ \bibnamefont {Koehl}},
  \bibinfo {author} {\bibfnamefont {J.~B.}\ \bibnamefont {Varley}}, \bibinfo
  {author} {\bibfnamefont {A.}~\bibnamefont {Janotti}}, \bibinfo {author}
  {\bibfnamefont {B.~B.}\ \bibnamefont {Buckley}}, \bibinfo {author}
  {\bibfnamefont {C.~G.}\ \bibnamefont {Van~de Walle}},\ and\ \bibinfo {author}
  {\bibfnamefont {D.~D.}\ \bibnamefont {Awschalom}},\ }\bibfield  {title}
  {\enquote {\bibinfo {title} {Quantum computing with defects},}\ }\href
  {https://doi.org/10.1073/pnas.1003052107} {\bibfield  {journal} {\bibinfo
  {journal} {Proc. Natl. Acad. Sci.}\ }\textbf {\bibinfo {volume} {107}},\
  \bibinfo {pages} {8513--8518} (\bibinfo {year} {2010})}\BibitemShut {NoStop}%
\bibitem [{\citenamefont {Gordon}\ \emph {et~al.}(2013)\citenamefont {Gordon},
  \citenamefont {Weber}, \citenamefont {Varley}, \citenamefont {Janotti},
  \citenamefont {Awschalom},\ and\ \citenamefont {Van~de
  Walle}}]{Gordon2013MRSBull}%
  \BibitemOpen
  \bibfield  {author} {\bibinfo {author} {\bibfnamefont {L.}~\bibnamefont
  {Gordon}}, \bibinfo {author} {\bibfnamefont {J.~R.}\ \bibnamefont {Weber}},
  \bibinfo {author} {\bibfnamefont {J.~B.}\ \bibnamefont {Varley}}, \bibinfo
  {author} {\bibfnamefont {A.}~\bibnamefont {Janotti}}, \bibinfo {author}
  {\bibfnamefont {D.~D.}\ \bibnamefont {Awschalom}},\ and\ \bibinfo {author}
  {\bibfnamefont {C.~G.}\ \bibnamefont {Van~de Walle}},\ }\bibfield  {title}
  {\enquote {\bibinfo {title} {Quantum computing with defects},}\ }\href
  {https://doi.org/10.1557/mrs.2013.206} {\bibfield  {journal} {\bibinfo
  {journal} {MRS Bull.}\ }\textbf {\bibinfo {volume} {38}},\ \bibinfo {pages}
  {802–807} (\bibinfo {year} {2013})}\BibitemShut {NoStop}%
\bibitem [{\citenamefont {Stewart}\ \emph {et~al.}(2019)\citenamefont
  {Stewart}, \citenamefont {Kianinia}, \citenamefont {Previdi}, \citenamefont
  {Tran}, \citenamefont {Aharonovich},\ and\ \citenamefont
  {Bradac}}]{Stewart2019OL}%
  \BibitemOpen
  \bibfield  {author} {\bibinfo {author} {\bibfnamefont {C.}~\bibnamefont
  {Stewart}}, \bibinfo {author} {\bibfnamefont {M.}~\bibnamefont {Kianinia}},
  \bibinfo {author} {\bibfnamefont {R.}~\bibnamefont {Previdi}}, \bibinfo
  {author} {\bibfnamefont {T.~T.}\ \bibnamefont {Tran}}, \bibinfo {author}
  {\bibfnamefont {I.}~\bibnamefont {Aharonovich}},\ and\ \bibinfo {author}
  {\bibfnamefont {C.}~\bibnamefont {Bradac}},\ }\bibfield  {title} {\enquote
  {\bibinfo {title} {Quantum emission from localized defects in zinc
  sulfide},}\ }\href {https://doi.org/10.1364/OL.44.004873} {\bibfield
  {journal} {\bibinfo  {journal} {Opt. Lett.}\ }\textbf {\bibinfo {volume}
  {44}},\ \bibinfo {pages} {4873--4876} (\bibinfo {year} {2019})}\BibitemShut
  {NoStop}%
\bibitem [{\citenamefont {Thiel}, \citenamefont {Böttger},\ and\ \citenamefont
  {Cone}(2011)}]{Thiel2011JL}%
  \BibitemOpen
  \bibfield  {author} {\bibinfo {author} {\bibfnamefont {C.}~\bibnamefont
  {Thiel}}, \bibinfo {author} {\bibfnamefont {T.}~\bibnamefont {Böttger}},\
  and\ \bibinfo {author} {\bibfnamefont {R.}~\bibnamefont {Cone}},\ }\bibfield
  {title} {\enquote {\bibinfo {title} {Rare-earth-doped materials for
  applications in quantum information storage and signal processing},}\ }\href
  {https://doi.org/https://doi.org/10.1016/j.jlumin.2010.12.015} {\bibfield
  {journal} {\bibinfo  {journal} {J. Lumin.}\ }\textbf {\bibinfo {volume}
  {131}},\ \bibinfo {pages} {353--361} (\bibinfo {year} {2011})}\BibitemShut
  {NoStop}%
\bibitem [{\citenamefont {Zhong}\ and\ \citenamefont
  {Goldner}(2019)}]{Zhong2019NP}%
  \BibitemOpen
  \bibfield  {author} {\bibinfo {author} {\bibfnamefont {T.}~\bibnamefont
  {Zhong}}\ and\ \bibinfo {author} {\bibfnamefont {P.}~\bibnamefont
  {Goldner}},\ }\bibfield  {title} {\enquote {\bibinfo {title} {Emerging
  rare-earth doped material platforms for quantum nanophotonics},}\ }\href
  {https://doi.org/doi:10.1515/nanoph-2019-0185} {\bibfield  {journal}
  {\bibinfo  {journal} {Nanophotonics}\ }\textbf {\bibinfo {volume} {8}},\
  \bibinfo {pages} {2003--2015} (\bibinfo {year} {2019})}\BibitemShut {NoStop}%
\bibitem [{\citenamefont {Mitchell}\ \emph {et~al.}(2021)\citenamefont
  {Mitchell}, \citenamefont {Austin}, \citenamefont {Timmerman}, \citenamefont
  {Dierolf},\ and\ \citenamefont {Fujiwara}}]{Mitchell2021NP}%
  \BibitemOpen
  \bibfield  {author} {\bibinfo {author} {\bibfnamefont {B.}~\bibnamefont
  {Mitchell}}, \bibinfo {author} {\bibfnamefont {H.}~\bibnamefont {Austin}},
  \bibinfo {author} {\bibfnamefont {D.}~\bibnamefont {Timmerman}}, \bibinfo
  {author} {\bibfnamefont {V.}~\bibnamefont {Dierolf}},\ and\ \bibinfo {author}
  {\bibfnamefont {Y.}~\bibnamefont {Fujiwara}},\ }\bibfield  {title} {\enquote
  {\bibinfo {title} {Temporally modulated energy shuffling in highly
  interconnected nanosystems},}\ }\href
  {https://doi.org/doi:10.1515/nanoph-2020-0484} {\bibfield  {journal}
  {\bibinfo  {journal} {Nanophotonics}\ }\textbf {\bibinfo {volume} {10}},\
  \bibinfo {pages} {851--876} (\bibinfo {year} {2021})}\BibitemShut {NoStop}%
\bibitem [{\citenamefont {Kinos}\ \emph {et~al.}(2021)\citenamefont {Kinos},
  \citenamefont {Hunger}, \citenamefont {Kolesov}, \citenamefont {Mølmer},
  \citenamefont {de~Riedmatten}, \citenamefont {Goldner}, \citenamefont
  {Tallaire}, \citenamefont {Morvan}, \citenamefont {Berger}, \citenamefont
  {Welinski}, \citenamefont {Karrai}, \citenamefont {Rippe}, \citenamefont
  {Kröll},\ and\ \citenamefont {Walther}}]{kinos2021roadmap}%
  \BibitemOpen
  \bibfield  {author} {\bibinfo {author} {\bibfnamefont {A.}~\bibnamefont
  {Kinos}}, \bibinfo {author} {\bibfnamefont {D.}~\bibnamefont {Hunger}},
  \bibinfo {author} {\bibfnamefont {R.}~\bibnamefont {Kolesov}}, \bibinfo
  {author} {\bibfnamefont {K.}~\bibnamefont {Mølmer}}, \bibinfo {author}
  {\bibfnamefont {H.}~\bibnamefont {de~Riedmatten}}, \bibinfo {author}
  {\bibfnamefont {P.}~\bibnamefont {Goldner}}, \bibinfo {author} {\bibfnamefont
  {A.}~\bibnamefont {Tallaire}}, \bibinfo {author} {\bibfnamefont
  {L.}~\bibnamefont {Morvan}}, \bibinfo {author} {\bibfnamefont
  {P.}~\bibnamefont {Berger}}, \bibinfo {author} {\bibfnamefont
  {S.}~\bibnamefont {Welinski}}, \bibinfo {author} {\bibfnamefont
  {K.}~\bibnamefont {Karrai}}, \bibinfo {author} {\bibfnamefont
  {L.}~\bibnamefont {Rippe}}, \bibinfo {author} {\bibfnamefont
  {S.}~\bibnamefont {Kröll}},\ and\ \bibinfo {author} {\bibfnamefont
  {A.}~\bibnamefont {Walther}},\ }\href@noop {} {\enquote {\bibinfo {title}
  {{Roadmap for Rare-earth Quantum Computing}},}\ } (\bibinfo {year} {2021}),\
  \Eprint {https://arxiv.org/abs/2103.15743} {arXiv:2103.15743 [quant-ph]}
  \BibitemShut {NoStop}%
\bibitem [{\citenamefont {Hommel}\ \emph {et~al.}(1985)\citenamefont {Hommel},
  \citenamefont {Hartmann}, \citenamefont {Godlewski}, \citenamefont {Langer},\
  and\ \citenamefont {Stapor}}]{Hommel1985JCG}%
  \BibitemOpen
  \bibfield  {author} {\bibinfo {author} {\bibfnamefont {D.}~\bibnamefont
  {Hommel}}, \bibinfo {author} {\bibfnamefont {H.}~\bibnamefont {Hartmann}},
  \bibinfo {author} {\bibfnamefont {M.}~\bibnamefont {Godlewski}}, \bibinfo
  {author} {\bibfnamefont {J.}~\bibnamefont {Langer}},\ and\ \bibinfo {author}
  {\bibfnamefont {A.}~\bibnamefont {Stapor}},\ }\bibfield  {title} {\enquote
  {\bibinfo {title} {Energy structure and recombination for {ZnS} bulk crystals
  doped with {Tb}, {Er} and {Eu}},}\ }\href
  {https://doi.org/https://doi.org/10.1016/0022-0248(85)90170-8} {\bibfield
  {journal} {\bibinfo  {journal} {J. Cryst. Growth}\ }\textbf {\bibinfo
  {volume} {72}},\ \bibinfo {pages} {346--350} (\bibinfo {year}
  {1985})}\BibitemShut {NoStop}%
\bibitem [{\citenamefont {Godlewski}\ and\ \citenamefont
  {Hommel}(1986)}]{Godlewski1986PSS}%
  \BibitemOpen
  \bibfield  {author} {\bibinfo {author} {\bibfnamefont {M.}~\bibnamefont
  {Godlewski}}\ and\ \bibinfo {author} {\bibfnamefont {D.}~\bibnamefont
  {Hommel}},\ }\bibfield  {title} {\enquote {\bibinfo {title} {Eu$^{2+}$
  photocharge transfer processes in {ZnS} crystals determined by photo-{ESR}
  measurements},}\ }\href
  {https://doi.org/https://doi.org/10.1002/pssa.2210950133} {\bibfield
  {journal} {\bibinfo  {journal} {phys. stat. solidi (a)}\ }\textbf {\bibinfo
  {volume} {95}},\ \bibinfo {pages} {261--268} (\bibinfo {year}
  {1986})}\BibitemShut {NoStop}%
\bibitem [{\citenamefont {Ehrhart}\ \emph {et~al.}(2008)\citenamefont
  {Ehrhart}, \citenamefont {Capoen}, \citenamefont {Robbe}, \citenamefont
  {Beclin}, \citenamefont {Boy}, \citenamefont {Turrell},\ and\ \citenamefont
  {Bouazaoui}}]{Ehrhart2008OM}%
  \BibitemOpen
  \bibfield  {author} {\bibinfo {author} {\bibfnamefont {G.}~\bibnamefont
  {Ehrhart}}, \bibinfo {author} {\bibfnamefont {B.}~\bibnamefont {Capoen}},
  \bibinfo {author} {\bibfnamefont {O.}~\bibnamefont {Robbe}}, \bibinfo
  {author} {\bibfnamefont {F.}~\bibnamefont {Beclin}}, \bibinfo {author}
  {\bibfnamefont {P.}~\bibnamefont {Boy}}, \bibinfo {author} {\bibfnamefont
  {S.}~\bibnamefont {Turrell}},\ and\ \bibinfo {author} {\bibfnamefont
  {M.}~\bibnamefont {Bouazaoui}},\ }\bibfield  {title} {\enquote {\bibinfo
  {title} {Energy transfer between semiconductor nanoparticles ({ZnS} or {CdS})
  and {Eu$^{3+}$} ions in sol–gel derived {ZrO$_2$} thin films},}\ }\href
  {https://doi.org/https://doi.org/10.1016/j.optmat.2007.10.004} {\bibfield
  {journal} {\bibinfo  {journal} {Opt. Mater.}\ }\textbf {\bibinfo {volume}
  {30}},\ \bibinfo {pages} {1595--1602} (\bibinfo {year} {2008})}\BibitemShut
  {NoStop}%
\bibitem [{\citenamefont {Wang}\ \emph {et~al.}(2015)\citenamefont {Wang},
  \citenamefont {Liang}, \citenamefont {Liu}, \citenamefont {Hu},\ and\
  \citenamefont {Fan}}]{Wang2015Nano}%
  \BibitemOpen
  \bibfield  {author} {\bibinfo {author} {\bibfnamefont {Y.}~\bibnamefont
  {Wang}}, \bibinfo {author} {\bibfnamefont {X.}~\bibnamefont {Liang}},
  \bibinfo {author} {\bibfnamefont {E.}~\bibnamefont {Liu}}, \bibinfo {author}
  {\bibfnamefont {X.}~\bibnamefont {Hu}},\ and\ \bibinfo {author}
  {\bibfnamefont {J.}~\bibnamefont {Fan}},\ }\bibfield  {title} {\enquote
  {\bibinfo {title} {Incorporation of lanthanide ({Eu$^{3+}$}) ions in {ZnS}
  semiconductor quantum dots with a trapped-dopant model and their
  photoluminescence spectroscopy study},}\ }\href
  {https://doi.org/10.1088/0957-4484/26/37/375601} {\bibfield  {journal}
  {\bibinfo  {journal} {Nanotechnology}\ }\textbf {\bibinfo {volume} {26}},\
  \bibinfo {pages} {375601} (\bibinfo {year} {2015})}\BibitemShut {NoStop}%
\bibitem [{\citenamefont {Horoz}\ \emph {et~al.}(2016)\citenamefont {Horoz},
  \citenamefont {Yakami}, \citenamefont {Poudyal}, \citenamefont {Pikal},
  \citenamefont {Wang},\ and\ \citenamefont {Tang}}]{Horoz2016AIPA}%
  \BibitemOpen
  \bibfield  {author} {\bibinfo {author} {\bibfnamefont {S.}~\bibnamefont
  {Horoz}}, \bibinfo {author} {\bibfnamefont {B.}~\bibnamefont {Yakami}},
  \bibinfo {author} {\bibfnamefont {U.}~\bibnamefont {Poudyal}}, \bibinfo
  {author} {\bibfnamefont {J.~M.}\ \bibnamefont {Pikal}}, \bibinfo {author}
  {\bibfnamefont {W.}~\bibnamefont {Wang}},\ and\ \bibinfo {author}
  {\bibfnamefont {J.}~\bibnamefont {Tang}},\ }\bibfield  {title} {\enquote
  {\bibinfo {title} {Controlled synthesis of {Eu$^{2+}$ and Eu$^{3+}$ doped
  ZnS} quantum dots and their photovoltaic and magnetic properties},}\ }\href
  {https://doi.org/10.1063/1.4948510} {\bibfield  {journal} {\bibinfo
  {journal} {AIP Advances}\ }\textbf {\bibinfo {volume} {6}},\ \bibinfo {pages}
  {045119} (\bibinfo {year} {2016})}\BibitemShut {NoStop}%
\bibitem [{\citenamefont {Wei}\ \emph {et~al.}(2018)\citenamefont {Wei},
  \citenamefont {Lu}, \citenamefont {Zhao}, \citenamefont {Zhao}, \citenamefont
  {Wang}, \citenamefont {Fu}, \citenamefont {Li}, \citenamefont {Guan},\ and\
  \citenamefont {Teng}}]{Wei2018ACSO}%
  \BibitemOpen
  \bibfield  {author} {\bibinfo {author} {\bibfnamefont {Z.}~\bibnamefont
  {Wei}}, \bibinfo {author} {\bibfnamefont {Y.}~\bibnamefont {Lu}}, \bibinfo
  {author} {\bibfnamefont {J.}~\bibnamefont {Zhao}}, \bibinfo {author}
  {\bibfnamefont {S.}~\bibnamefont {Zhao}}, \bibinfo {author} {\bibfnamefont
  {R.}~\bibnamefont {Wang}}, \bibinfo {author} {\bibfnamefont {N.}~\bibnamefont
  {Fu}}, \bibinfo {author} {\bibfnamefont {X.}~\bibnamefont {Li}}, \bibinfo
  {author} {\bibfnamefont {L.}~\bibnamefont {Guan}},\ and\ \bibinfo {author}
  {\bibfnamefont {F.}~\bibnamefont {Teng}},\ }\bibfield  {title} {\enquote
  {\bibinfo {title} {{Synthesis and Luminescent Modulation of ZnS Crystallite
  by a Hydrothermal Method}},}\ }\href
  {https://doi.org/10.1021/acsomega.7b01574} {\bibfield  {journal} {\bibinfo
  {journal} {ACS Omega}\ }\textbf {\bibinfo {volume} {3}},\ \bibinfo {pages}
  {137--143} (\bibinfo {year} {2018})}\BibitemShut {NoStop}%
\bibitem [{\citenamefont {Watts}\ and\ \citenamefont
  {Holton}(1968)}]{Watts1968PR}%
  \BibitemOpen
  \bibfield  {author} {\bibinfo {author} {\bibfnamefont {R.~K.}\ \bibnamefont
  {Watts}}\ and\ \bibinfo {author} {\bibfnamefont {W.~C.}\ \bibnamefont
  {Holton}},\ }\bibfield  {title} {\enquote {\bibinfo {title}
  {{Paramagnetic-Resonance Studies of Rare-Earth Impurities in II-VI
  Compounds}},}\ }\href {https://doi.org/10.1103/PhysRev.173.417} {\bibfield
  {journal} {\bibinfo  {journal} {Phys. Rev.}\ }\textbf {\bibinfo {volume}
  {173}},\ \bibinfo {pages} {417--426} (\bibinfo {year} {1968})}\BibitemShut
  {NoStop}%
\bibitem [{\citenamefont {Varley}\ and\ \citenamefont
  {Lordi}(2013)}]{Varley2013APL}%
  \BibitemOpen
  \bibfield  {author} {\bibinfo {author} {\bibfnamefont {J.~B.}\ \bibnamefont
  {Varley}}\ and\ \bibinfo {author} {\bibfnamefont {V.}~\bibnamefont {Lordi}},\
  }\bibfield  {title} {\enquote {\bibinfo {title} {{Electrical properties of
  point defects in CdS and ZnS}},}\ }\href {https://doi.org/10.1063/1.4819492}
  {\bibfield  {journal} {\bibinfo  {journal} {Appl. Phys. Lett.}\ }\textbf
  {\bibinfo {volume} {103}},\ \bibinfo {pages} {102103} (\bibinfo {year}
  {2013})}\BibitemShut {NoStop}%
\bibitem [{\citenamefont {Varley}\ and\ \citenamefont
  {Lordi}(2014)}]{Varley2014JAP}%
  \BibitemOpen
  \bibfield  {author} {\bibinfo {author} {\bibfnamefont {J.~B.}\ \bibnamefont
  {Varley}}\ and\ \bibinfo {author} {\bibfnamefont {V.}~\bibnamefont {Lordi}},\
  }\bibfield  {title} {\enquote {\bibinfo {title} {Intermixing at the
  absorber-buffer layer interface in thin-film solar cells: {The electronic
  effects of point defects in Cu(In,Ga)(Se,S)$_2$ and Cu$_2$ZnSn(Se,S)$_4$
  devices}},}\ }\href {https://doi.org/10.1063/1.4892407} {\bibfield  {journal}
  {\bibinfo  {journal} {J. Appl. Phys.}\ }\textbf {\bibinfo {volume} {116}},\
  \bibinfo {pages} {063505} (\bibinfo {year} {2014})}\BibitemShut {NoStop}%
\bibitem [{\citenamefont {Hoang}, \citenamefont {Latouche},\ and\ \citenamefont
  {Jobic}(2019)}]{Hoang2019CMS}%
  \BibitemOpen
  \bibfield  {author} {\bibinfo {author} {\bibfnamefont {K.}~\bibnamefont
  {Hoang}}, \bibinfo {author} {\bibfnamefont {C.}~\bibnamefont {Latouche}},\
  and\ \bibinfo {author} {\bibfnamefont {S.}~\bibnamefont {Jobic}},\ }\bibfield
   {title} {\enquote {\bibinfo {title} {{Defect energy levels and persistent
  luminescence in Cu-doped ZnS}},}\ }\href
  {https://doi.org/https://doi.org/10.1016/j.commatsci.2019.03.016} {\bibfield
  {journal} {\bibinfo  {journal} {Comput. Mater. Sci.}\ }\textbf {\bibinfo
  {volume} {163}},\ \bibinfo {pages} {63--67} (\bibinfo {year}
  {2019})}\BibitemShut {NoStop}%
\bibitem [{\citenamefont {Hoang}(2021)}]{Hoang2021PRM}%
  \BibitemOpen
  \bibfield  {author} {\bibinfo {author} {\bibfnamefont {K.}~\bibnamefont
  {Hoang}},\ }\bibfield  {title} {\enquote {\bibinfo {title} {Tuning the
  valence and concentration of europium and luminescence centers in {GaN}
  through co-doping and defect association},}\ }\href
  {https://doi.org/10.1103/PhysRevMaterials.5.034601} {\bibfield  {journal}
  {\bibinfo  {journal} {Phys. Rev. Materials}\ }\textbf {\bibinfo {volume}
  {5}},\ \bibinfo {pages} {034601} (\bibinfo {year} {2021})}\BibitemShut
  {NoStop}%
\bibitem [{\citenamefont {Freysoldt}\ \emph {et~al.}(2014)\citenamefont
  {Freysoldt}, \citenamefont {Grabowski}, \citenamefont {Hickel}, \citenamefont
  {Neugebauer}, \citenamefont {Kresse}, \citenamefont {Janotti},\ and\
  \citenamefont {{Van de Walle}}}]{Freysoldt2014RMP}%
  \BibitemOpen
  \bibfield  {author} {\bibinfo {author} {\bibfnamefont {C.}~\bibnamefont
  {Freysoldt}}, \bibinfo {author} {\bibfnamefont {B.}~\bibnamefont
  {Grabowski}}, \bibinfo {author} {\bibfnamefont {T.}~\bibnamefont {Hickel}},
  \bibinfo {author} {\bibfnamefont {J.}~\bibnamefont {Neugebauer}}, \bibinfo
  {author} {\bibfnamefont {G.}~\bibnamefont {Kresse}}, \bibinfo {author}
  {\bibfnamefont {A.}~\bibnamefont {Janotti}},\ and\ \bibinfo {author}
  {\bibfnamefont {C.~G.}\ \bibnamefont {{Van de Walle}}},\ }\bibfield  {title}
  {\enquote {\bibinfo {title} {First-principles calculations for point defects
  in solids},}\ }\href {https://doi.org/10.1103/RevModPhys.86.253} {\bibfield
  {journal} {\bibinfo  {journal} {Rev. Mod. Phys.}\ }\textbf {\bibinfo {volume}
  {86}},\ \bibinfo {pages} {253--305} (\bibinfo {year} {2014})}\BibitemShut
  {NoStop}%
\bibitem [{\citenamefont {Freysoldt}, \citenamefont {Neugebauer},\ and\
  \citenamefont {{Van de Walle}}(2009)}]{Freysoldt}%
  \BibitemOpen
  \bibfield  {author} {\bibinfo {author} {\bibfnamefont {C.}~\bibnamefont
  {Freysoldt}}, \bibinfo {author} {\bibfnamefont {J.}~\bibnamefont
  {Neugebauer}},\ and\ \bibinfo {author} {\bibfnamefont {C.~G.}\ \bibnamefont
  {{Van de Walle}}},\ }\bibfield  {title} {\enquote {\bibinfo {title} {{Fully
  \textit{Ab Initio} Finite-Size Corrections for Charged-Defect Supercell
  Calculations}},}\ }\href {https://doi.org/10.1103/PhysRevLett.102.016402}
  {\bibfield  {journal} {\bibinfo  {journal} {Phys. Rev. Lett.}\ }\textbf
  {\bibinfo {volume} {102}},\ \bibinfo {pages} {016402} (\bibinfo {year}
  {2009})}\BibitemShut {NoStop}%
\bibitem [{\citenamefont {Freysoldt}, \citenamefont {Neugebauer},\ and\
  \citenamefont {{Van de Walle}}(2011)}]{Freysoldt11}%
  \BibitemOpen
  \bibfield  {author} {\bibinfo {author} {\bibfnamefont {C.}~\bibnamefont
  {Freysoldt}}, \bibinfo {author} {\bibfnamefont {J.}~\bibnamefont
  {Neugebauer}},\ and\ \bibinfo {author} {\bibfnamefont {C.~G.}\ \bibnamefont
  {{Van de Walle}}},\ }\bibfield  {title} {\enquote {\bibinfo {title}
  {Electrostatic interactions between charged defects in supercells},}\ }\href
  {https://doi.org/10.1002/pssb.201046289} {\bibfield  {journal} {\bibinfo
  {journal} {phys. status solidi (b)}\ }\textbf {\bibinfo {volume} {248}},\
  \bibinfo {pages} {1067--1076} (\bibinfo {year} {2011})}\BibitemShut {NoStop}%
\bibitem [{\citenamefont {{Van de Walle}}\ and\ \citenamefont
  {Neugebauer}(2004)}]{walle:3851}%
  \BibitemOpen
  \bibfield  {author} {\bibinfo {author} {\bibfnamefont {C.~G.}\ \bibnamefont
  {{Van de Walle}}}\ and\ \bibinfo {author} {\bibfnamefont {J.}~\bibnamefont
  {Neugebauer}},\ }\bibfield  {title} {\enquote {\bibinfo {title}
  {{First-principles calculations for defects and impurities: Applications to
  III-nitrides}},}\ }\href {https://doi.org/10.1063/1.1682673} {\bibfield
  {journal} {\bibinfo  {journal} {J. Appl. Phys.}\ }\textbf {\bibinfo {volume}
  {95}},\ \bibinfo {pages} {3851--3879} (\bibinfo {year} {2004})}\BibitemShut
  {NoStop}%
\bibitem [{\citenamefont {Hoang}\ and\ \citenamefont
  {Johannes}(2018)}]{Hoang2018JPCM}%
  \BibitemOpen
  \bibfield  {author} {\bibinfo {author} {\bibfnamefont {K.}~\bibnamefont
  {Hoang}}\ and\ \bibinfo {author} {\bibfnamefont {M.~D.}\ \bibnamefont
  {Johannes}},\ }\bibfield  {title} {\enquote {\bibinfo {title} {Defect physics
  in complex energy materials},}\ }\href
  {https://doi.org/10.1088/1361-648x/aacb05} {\bibfield  {journal} {\bibinfo
  {journal} {J. Phys.: Condens. Matter}\ }\textbf {\bibinfo {volume} {30}},\
  \bibinfo {pages} {293001} (\bibinfo {year} {2018})}\BibitemShut {NoStop}%
\bibitem [{\citenamefont {Heyd}, \citenamefont {Scuseria},\ and\ \citenamefont
  {Ernzerhof}(2003)}]{heyd:8207}%
  \BibitemOpen
  \bibfield  {author} {\bibinfo {author} {\bibfnamefont {J.}~\bibnamefont
  {Heyd}}, \bibinfo {author} {\bibfnamefont {G.~E.}\ \bibnamefont {Scuseria}},\
  and\ \bibinfo {author} {\bibfnamefont {M.}~\bibnamefont {Ernzerhof}},\
  }\bibfield  {title} {\enquote {\bibinfo {title} {{Hybrid functionals based on
  a screened Coulomb potential}},}\ }\href {https://doi.org/10.1063/1.1564060}
  {\bibfield  {journal} {\bibinfo  {journal} {J. Chem. Phys.}\ }\textbf
  {\bibinfo {volume} {118}},\ \bibinfo {pages} {8207--8215} (\bibinfo {year}
  {2003})}\BibitemShut {NoStop}%
\bibitem [{\citenamefont {Bl\"ochl}(1994)}]{PAW1}%
  \BibitemOpen
  \bibfield  {author} {\bibinfo {author} {\bibfnamefont {P.~E.}\ \bibnamefont
  {Bl\"ochl}},\ }\bibfield  {title} {\enquote {\bibinfo {title} {Projector
  augmented-wave method},}\ }\href {https://doi.org/10.1103/PhysRevB.50.17953}
  {\bibfield  {journal} {\bibinfo  {journal} {Phys. Rev. B}\ }\textbf {\bibinfo
  {volume} {50}},\ \bibinfo {pages} {17953--17979} (\bibinfo {year}
  {1994})}\BibitemShut {NoStop}%
\bibitem [{\citenamefont {Kresse}\ and\ \citenamefont
  {Furthm\"uller}(1996)}]{VASP2}%
  \BibitemOpen
  \bibfield  {author} {\bibinfo {author} {\bibfnamefont {G.}~\bibnamefont
  {Kresse}}\ and\ \bibinfo {author} {\bibfnamefont {J.}~\bibnamefont
  {Furthm\"uller}},\ }\bibfield  {title} {\enquote {\bibinfo {title} {Efficient
  iterative schemes for ab initio total-energy calculations using a plane-wave
  basis set},}\ }\href {https://doi.org/10.1103/PhysRevB.54.11169} {\bibfield
  {journal} {\bibinfo  {journal} {Phys. Rev. B}\ }\textbf {\bibinfo {volume}
  {54}},\ \bibinfo {pages} {11169--11186} (\bibinfo {year} {1996})}\BibitemShut
  {NoStop}%
\bibitem [{\citenamefont {Dudarev}\ \emph {et~al.}(1998)\citenamefont
  {Dudarev}, \citenamefont {Botton}, \citenamefont {Savrasov}, \citenamefont
  {Humphreys},\ and\ \citenamefont {Sutton}}]{Dudarev1998}%
  \BibitemOpen
  \bibfield  {author} {\bibinfo {author} {\bibfnamefont {S.~L.}\ \bibnamefont
  {Dudarev}}, \bibinfo {author} {\bibfnamefont {G.~A.}\ \bibnamefont {Botton}},
  \bibinfo {author} {\bibfnamefont {S.~Y.}\ \bibnamefont {Savrasov}}, \bibinfo
  {author} {\bibfnamefont {C.~J.}\ \bibnamefont {Humphreys}},\ and\ \bibinfo
  {author} {\bibfnamefont {A.~P.}\ \bibnamefont {Sutton}},\ }\bibfield  {title}
  {\enquote {\bibinfo {title} {{Electron-energy-loss spectra and the structural
  stability of nickel oxide: An LSDA$+$$U$ study}},}\ }\href
  {https://doi.org/10.1103/PhysRevB.57.1505} {\bibfield  {journal} {\bibinfo
  {journal} {Phys. Rev. B}\ }\textbf {\bibinfo {volume} {57}},\ \bibinfo
  {pages} {1505--1509} (\bibinfo {year} {1998})}\BibitemShut {NoStop}%
\bibitem [{\citenamefont {Perdew}, \citenamefont {Burke},\ and\ \citenamefont
  {Ernzerhof}(1996)}]{GGA}%
  \BibitemOpen
  \bibfield  {author} {\bibinfo {author} {\bibfnamefont {J.~P.}\ \bibnamefont
  {Perdew}}, \bibinfo {author} {\bibfnamefont {K.}~\bibnamefont {Burke}},\ and\
  \bibinfo {author} {\bibfnamefont {M.}~\bibnamefont {Ernzerhof}},\ }\bibfield
  {title} {\enquote {\bibinfo {title} {Generalized gradient approximation made
  simple},}\ }\href {https://doi.org/10.1103/PhysRevLett.77.3865} {\bibfield
  {journal} {\bibinfo  {journal} {Phys. Rev. Lett.}\ }\textbf {\bibinfo
  {volume} {77}},\ \bibinfo {pages} {3865--3868} (\bibinfo {year}
  {1996})}\BibitemShut {NoStop}%
\bibitem [{\citenamefont {Buckeridge}(2019)}]{Buckeridge2019CPC}%
  \BibitemOpen
  \bibfield  {author} {\bibinfo {author} {\bibfnamefont {J.}~\bibnamefont
  {Buckeridge}},\ }\bibfield  {title} {\enquote {\bibinfo {title} {{Equilibrium
  point defect and charge carrier concentrations in a material determined
  through calculation of the self-consistent Fermi energy}},}\ }\href
  {https://doi.org/10.1016/j.cpc.2019.06.017} {\bibfield  {journal} {\bibinfo
  {journal} {Comput. Phys. Commun.}\ }\textbf {\bibinfo {volume} {244}},\
  \bibinfo {pages} {329--342} (\bibinfo {year} {2019})}\BibitemShut {NoStop}%
\bibitem [{\citenamefont {Lee}, \citenamefont {O'Donnell},\ and\ \citenamefont
  {Watkins}(1982)}]{Lee1982SSC}%
  \BibitemOpen
  \bibfield  {author} {\bibinfo {author} {\bibfnamefont {K.}~\bibnamefont
  {Lee}}, \bibinfo {author} {\bibfnamefont {K.}~\bibnamefont {O'Donnell}},\
  and\ \bibinfo {author} {\bibfnamefont {G.}~\bibnamefont {Watkins}},\
  }\bibfield  {title} {\enquote {\bibinfo {title} {Optically detected magnetic
  resonance of the zinc vacancy in {ZnS}},}\ }\href
  {https://doi.org/https://doi.org/10.1016/0038-1098(82)91228-5} {\bibfield
  {journal} {\bibinfo  {journal} {Solid State Commun.}\ }\textbf {\bibinfo
  {volume} {41}},\ \bibinfo {pages} {881--883} (\bibinfo {year}
  {1982})}\BibitemShut {NoStop}%
\end{thebibliography}
%

\end{document}